\documentclass{emulateapj}
\slugcomment{{\sc Accepted to ApJ:} September 29, 2011}

\newcommand{\Teff}{T_{\mathrm{eff}}}
\def \astrosun {\mbox{$\odot$}}
\newcommand{\Msun}{\ensuremath{\mbox{M}_{\astrosun}}}
\newcommand{\Mjup}{\ensuremath{\mbox{M}_{\mathrm{Jup}}}}
\newcommand{\kms}{\mathrm{km \, s^{-1}}}

\usepackage{amsmath, amsfonts, amssymb, color, verbatim, mathrsfs, natbib}
\begin{document}

\shorttitle{No Planet Around HD 149382}
\title{Non-Detection of the Putative Substellar Companion to HD 149382}

\author{Jackson M. Norris\altaffilmark{1}, 
Jason T. Wright\altaffilmark{1}, 
Richard A. Wade, 
Suvrath Mahadevan\altaffilmark{1}, 
Sara Gettel\altaffilmark{1}}

\affil{Department of Astronomy and Astrophysics, 525 Davey Lab, The
  Pennsylvania State University, University Park, PA 16802}
\altaffiltext{1}{Center for Exoplanets and Habitable Worlds, 525 Davey
  Lab, The Pennsylvania State University, University Park, PA 16802}

\begin{abstract}

It has been argued that a substellar companion may significantly
influence the evolution of the progenitors of sdB stars.  Recently, the
bright sdB star HD 149382 has been claimed to host a substellar
(possibly planetary) companion with a period of 2.391 days.  This has
important implications for the evolution of the progenitors of sdB stars
as well as the source of the UV-excess seen in elliptical galaxies.  In
order to verify this putative planet, we made 10 radial velocity measurements of HD
149382 over 17 days with the High Resolution Spectrograph at the
Hobby-Eberly Telescope.  Our data conclusively demonstrate that the putative substellar
companion does not exist, and they exclude the presence of almost
any substellar companion with $P < 28$ days and $M\sin{i} \gtrsim 1 \, \Mjup$. \\

\end{abstract}


\section{Introduction} \label{sec:intro}        

The nature of the progenitors of subdwarf B (sdB) stars is not well understood. One
theoretical formation method is through a common-envelope (CE) phase of
evolution with an orbital companion \citep{han+2002,han+2003}. As the progenitor
star ascends the first red giant branch, interaction
with a companion via a CE may cause it to lose large
amounts of mass. When the progenitor star finally ignites core helium burning on
the horizontal branch, it has a small radius, a high effective
temperature, and a low-mass hydrogen envelope that
cannot sustain shell burning: it is a sdB star. \citet{soker1998} suggests
that a substellar companion (a
brown dwarf or planet) may be massive enough to survive this CE phase
without merging with the host star. This implies that some sdBs may be
created via the evolution of single stars with large planetary
companions.  In addition to their importance as tracers of the many
paths of stellar evolution, sdB stars have broader importance in that
they may be the dominant source of the UV-excess seen in elliptical
galaxies \citep{brown+1997,brown+2008}.

HD 149382 is the brightest known sdB star, and thus provides an
excellent opportunity to better understand the progenitors of sdB stars. 
It is an sdOB star with atmospheric parameters $T_\mathrm{eff} = 35500 \pm 500$~K,
$\log g = 5.80 \pm 0.05$ \citep[hereinafter G09]{geier+2009}, 
and a pattern of photospheric abundances altered by diffusion
processes \citep[][who found $T_\mathrm{eff} = 35000 \pm 2000$~K, $\log g
= 5.5 \pm 0.3$]{Baschek+82}. G09 claim to have detected a substellar
companion to HD 149382 with a mass of $8-23 \, \Mjup$ using 16 radial
velocity (RV) measurements collected from four different telescopes over
a span of four years. G09 claim this is significant because it suggests that the
progenitors of sdB stars can be stripped of their hydrogen envelopes by
substellar companions during the CE phase, strongly supporting
\citet{soker1998}.  This is the first claim of a close substellar
companion to an sdB star able to affect the evolution of its host.  For
brevity we will use the word ``planet'' to refer to a substellar
companion of HD 149382, whether a true planet or a brown dwarf.

\citet{jacobs+2011} reported a non-detection of the claimed planet
from their own RV measurements of HD 149382 using the HERMES spectrograph \citep{raskin+2011}. Their
upper limit on the RV variability of HD 149382 was $0.8 \, \kms$,
nearly three times lower than the amplitude reported by
G09.  Since a planet with the reported short orbital period would have
important implications, we independently sought verification of its
existence, as we describe here. 

\section{Data} \label{sec:data}          

We observed HD 149382 on 10 nights between April 10-27, 2010 using the
High Resolution Spectrograph \citep[HRS;][]{Tull98} in its $R=60,000$
configuration, at the Hobby-Eberly Telescope \citep[HET;][]{Ramsey98}, a
9.2m telescope stationed at the McDonald Observatory near Fort Davis,
Texas.  We exploited the queue scheduling of this telescope
\citep{Shetrone07} to efficiently make our 10 observations and acquire good
phase coverage of the orbit.  Typical single-exposure signal-to-noise
ratios at blaze peak near 5500 \AA\ were
100 per 4-pixel resolution element.  The use of a single, fiber-fed
spectrograph should reduce the potentially significant systematic
velocity errors that G09 may have incurred combining measurements made
from four different telescopes.

Our observing procedure was to obtain two exposures of the star through
the $2\arcsec$ fiber followed immediately by two exposures with the iodine
absorption cell inserted in the beam.  The cell served to measure
night-to-night instrumental offsets due to imperfect repeatability of
the instrument configuration.  We also obtained bracketing thorium-argon
(ThAr) lamp exposures through the calibration fiber, and the standard
ThAr exposures and flat lamp exposures through the iodine cell at the
beginning or end of each night.  On the first night, we obtained only one
stellar exposure of each type, and on the final night we obtained
three.

\subsection{Reducing Spectra} \label{sec:reducing_spectra}

The spectra were reduced using \texttt{REDUCE}, an advanced image
reduction package designed to reduce 2-dimensional, raw echelle spectra
into 1-dimensional, wavelength-calibrated spectra \citep{REDUCE}.  The
reduction was modified slightly to accommodate fiber-fed spectra, rather than slit
spectra.  Each observation contains information on two charge-coupled
devices (CCDs), a ``blue'' CCD and a
``red'' CCD; we limited our analysis to the blue CCD, where we expect
most velocity information to be located. We extracted 47
orders of 1-dimensional spectra from the blue CCD.

When determining velocities, we stacked all stellar exposures from the
same night into a single spectrum.  The time between stellar exposures on
the same night is always less than 31 minutes; finding RV from a stacked
spectrum for each night should not change our results, since any change
in the RV of the star will be small over the ``smoothing interval'' of
our exposures.  Since we combined pure stellar exposures with
stellar-plus-iodine-cell exposures, we were careful to analyze only those
orders that contain no iodine lines.

\subsection{Visual Companion} \label{sec:visual_companion}

In addition to the putative substellar companion, HD 149382 features a
known visual, stellar companion.  \citet{ostensen+2005} detected this
visual companion using adaptive optics.  The visual companion is about
$1\arcsec$ away \citep[corresponding to 75 AU,][]{jacobs+2011} while the HRS
fiber diameter is $2\arcsec$, so we likely collected light from both stars.
We therefore took care in our velocity analysis to select only lines
expected from the primary.  \citet{ulla+1998} detected an infrared
excess in HD 149382, probably due to this visual companion, but
\citet{geier+2010} detect no spectral features from the companion in the
optical.  Since we also observed in the optical, and since any orbital
motion between the stars must be too slow to be detectable over a 17 day
span, we do not expect significant RV errors due to the companion
spectrum.

\section{Method} \label{sec:method}        

\subsection{Line Selection} \label{sec:line_selection}

To aid in selecting absorption lines that are unequivocally from the primary star,
and not from the faint companion, we generated a synthetic spectrum from
a Castelli \& Kurucz model\footnote{See \texttt{http://kurucz.harvard.edu}} \citep{castelli+2003} with the approximate parameters of HD 149382, as reported
by G09.  We started from a solar-abundance Kurucz atmosphere model in
LTE with $\Teff = 35000 \, \mathrm{K}$, $\log g = 5.0$. 

We selected a line list from the Vienna Atomic Line Database
\citep[VALD;][]{Kupka00} spanning the wavelength range $415-500 \, \mathrm{nm}$, and
corresponding to the model selected.  We used the \texttt{SYNTH3} code
\citep{Kochukhov07} to generate a stellar spectrum, adopting abundances
for elements derived by \citet{Baschek+82}, and solar abundances of
\citet{Asplund09} for remaining elements. The resulting spectrum was
broadened with a rotational profile of ($v \sin{i} = 2.4 \, \kms$), and
convolved with a Gaussian profile to match the $R = 60,000$ HRS
observation.

The synthetic spectrum is not an exact match to the observed spectrum of
HD 149382, as several lines are present in the model but not the data (and
vice versa). We note though, that such an exact match is not required for
our purpose. The synthetic spectrum is of sufficient fidelity to
unambiguously identify lines that belong to the primary star, and we chose these
lines (i.e. present in the model {\em and} the data) to measure the RVs.

\begin{figure}\begin{center}   
\plotone{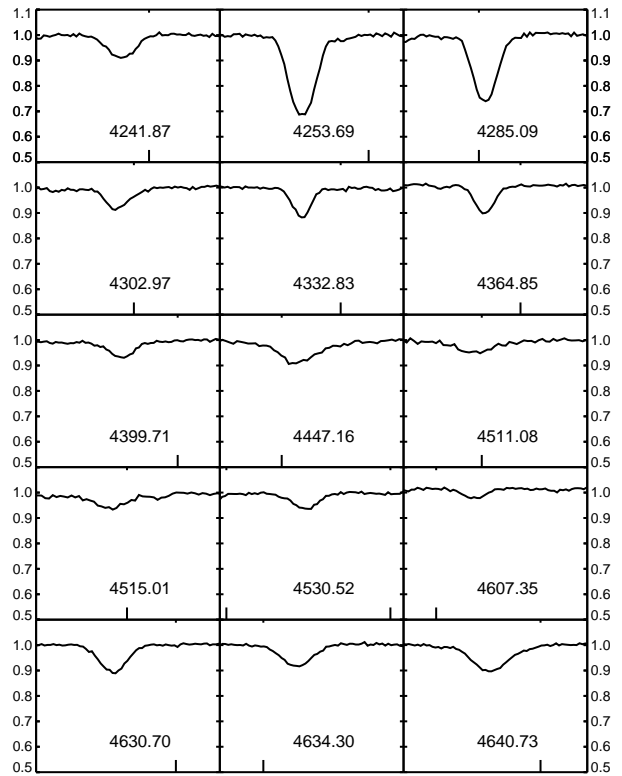}
\caption{\label{fig:lines} An atlas illustrating the 15 lines used in
  this RV analysis.  Each plot presents a region of the spectrum
  centered on the indicated waveglength (listed in \AA)
  averaged over all observations.  The span of the x-axis in each plot is given in
  Table~\ref{tab:line}.  The tick marks on the x-axes represent
  integer air wavelengths in \AA.} 
\end{center}\end{figure}

We visually selected 15 regions containing moderately deep and narrow 
metal absorption features. Each of these 15 pixel groups contains one or
more lines at wavelengths predicted from the model.  Our final selection
of absorption features is given in Table \ref{tab:line}, and the
averaged observed spectrum for each pixel group is shown in
Figure~\ref{fig:lines}.

\begin{deluxetable}{ccclc}  
\tablecolumns{5}
\tablecaption{Absorption Features Used for RV Measurement} 
\tablehead{
	\colhead{Min. $\lambda$} &
	\colhead{Max. $\lambda$} &
	\colhead{Median $\lambda$} &
	\colhead{Species} &
	\colhead{Lines} \\
	\colhead{[\AA] (air)} &
	\colhead{[\AA] (air)} &
	\colhead{[\AA] (air)} &
	\colhead{} &
	\colhead{}
}
\startdata

4241.33   &4242.42   &4241.87   &\ion{N}{2}                             &2 \\
4253.19   &4254.19   &4253.69   &\ion{Fe}{3}, \ion{S}{3}, \ion{O}{2}    &5 \\
4284.59   &4285.59   &4285.09   &\ion{Ca}{3}, \ion{S}{3}, \ion{O}{2}    &3 \\
4302.42   &4303.51   &4302.97   &\ion{Ca}{3}, \ion{O}{2}                &3 \\
4332.27   &4333.38   &4332.83   &\ion{S}{3}, \ion{O}{2}, \ion{N}{3}     &3 \\
4364.30   &4365.40   &4364.85   &\ion{S}{3}                             &1 \\
4399.16   &4400.25   &4399.71   &\ion{Ca}{3}, \ion{Zn}{3}               &2 \\
4446.67   &4447.65   &4447.16   &\ion{N}{2}                             &1 \\
4510.57   &4511.58   &4511.08   &\ion{N}{3}, \ion{Ne}{2}                &4 \\
4514.52   &4515.49   &4515.01   &\ion{N}{3}, \ion{Ne}{2}, \ion{C}{3}    &3 \\
4529.96   &4531.08   &4530.52   &\ion{N}{2}, \ion{N}{3}                 &2 \\
4606.81   &4607.88   &4607.35   &\ion{Ne}{2}, \ion{N}{2}                &2 \\
4630.11   &4631.28   &4630.70   &\ion{N}{2}, \ion{O}{3}, \ion{Si}{4}    &5 \\
4633.73   &4634.87   &4634.30   &\ion{N}{3}, \ion{Ne}{2}                &2 \\
4640.18   &4641.28   &4640.73   &\ion{N}{3}                             &1\\
\enddata
\tablecomments{The absorption features predicted from our synthetic
  spectrum in our 15 pixel groups.  Only the lines whose flux minima
  occur within the search area are included in the tabulation 
  of species and number of lines.  }
\label{tab:line}
\end{deluxetable}

\subsection{Radial Velocity Determination}

A potential source of systematic error is the stability of the
HRS wavelength solution.  We checked our ability to determine precise
wavelength solutions on different nights by comparing the night-to-night
shift in the wavelength solution to the shifts we measured in the iodine
region of the stellar spectra.  Since the iodine exposures were made
through the science fiber and have extremely high Doppler content, they
provide a highly sensitive measurement of any instrumental drift
affecting stellar velocity measurements.  We compared the shifts
measured in 192 63-pixel sections of the iodine region (through a
cross-correlation analysis in pixel space) to the shifts we
measured in the same sections using the nearest adjacent ThAr exposures
(made through the calibration fiber).
On each night, the median shift measured from all 192 sections differs by
$\lesssim 0.1$ pixel between the ThAr exposures and the
iodine region of the stellar spectra, indicating that the
ThAr exposures track the instrumental shifts of the basement-mounted,
temperature-controlled HRS as observed through the science fiber to better than $\sim 100 \, \mathrm{m \, s^{-1}}$.
The stability of the wavelength solution (i.e., its
night-to-night precision) is not a limiting factor in our RV precision.

To determine a RV for each night's average spectrum, we rebinned our
model spectrum into the wavelengths of each pixel group, using the
wavelength solution determined from the adjacent ThAr exposures.  We
then calculated the peak in the cross-correlation function (CCF) in
pixel space between our model and our observation, and used the
dispersion of the wavelength solution to calculate a velocity shift.
The primary source of uncertainty in our RV measurements is the
uncertainty in the location of the CCF peak due to the finite
signal-to-noise ratio of the data.

Each of the 15 pixel groups yielded its own set of RV measurements.
We subtracted the barycentric motion of the telescope using routines
provided by the California Planet Survey, which are proven below $0.001 \, \kms$ \citep[e.g.][]{Howard10a}.  Over the timespan of our data,
the change in barycentric correction amounts to almost $7 \, \kms$.
Probably because of line blends, the CCFs for the different regions
yield RVs that do not share a common apparent systemic velocity, although they
indepently detect the barycentric motion.  Since we are primarily concerned with the
acceleration of the star in its 
putative orbit rather than the systemic velocity of the star, we
subtracted the mean RV of each pixel group from its time series. For
each night, we then averaged the resulting differential RVs derived from
each pixel group.  We used the standard error of the mean of the relative RVs
to determine the ``internal'' error on each night's
measurement.  Our RVs thus have an arbitrary zero point.
We present our results after applying these corrections in
Table~\ref{tab:rv}.

\begin{deluxetable}{cc} 
\tablecolumns{2}
\tablecaption{Radial Velocities}
\tablehead{
	\colhead{Barycentric Julian Date} & 
	\colhead{Velocity} \\
	\colhead{[-2455000 days]} &
	\colhead{[km/s]}
}
\startdata
296.924			&-0.06$\, \pm \, 0.12$					\\
298.915			&\phantom{-}0.01$\, \pm\, 0.06$			\\
302.906			&-0.13$\, \pm\, 0.07$					\\
305.898			&-0.02$\, \pm\, 0.06$					\\
307.878			&\phantom{-}0.23$\, \pm\, 0.10$			\\
308.884			&\phantom{-}0.17$\, \pm\, 0.15$			\\
309.896			&-0.24$\, \pm\, 0.15$					\\
311.876			&\phantom{-}0.10$\, \pm\, 0.10$			\\
312.884			&-0.05$\, \pm\, 0.09$					\\
313.900			&\phantom{-}0.07$\, \pm\, 0.15$			\\
\enddata  
\label{tab:rv}
\end{deluxetable}

\section{Results} \label{sec:results}  

The root-mean-square variation of our measured velocities is $0.13 \, \kms$.
The (reduced) $\chi^2_\nu$ value for the null hypothesis of
no measured stellar acceleration is 1.61, suggesting that either our
uncertainties are slight underestimates or that the target is
intrinsically variable at this level.

To verify that these measurements are inconsistent with the claim of
G09, we fitted our RVs to a circular Keplerian model using the
\texttt{RVLIN} code \citep{Wright09b}, with an initial period estimate equal
to the value quoted by G09.  Because we are testing a specific
hypothesis, we limited the solutions to values within 5-sigma of the
values quoted by G09, specifically $P=\{2.381, 2.401 \}
\, \mathrm{days}$.  We allowed all other orbital elements in the fit to
float, including the phase, since the time elapsed since the G09
measurements is sufficiently long that the propagated phase is uncertain.

\begin{figure}               
\plotone{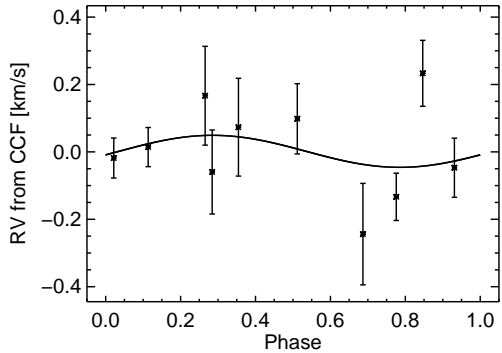} 
\caption{Best fitted phase-folded RV curve for HD 149382, limiting the period to a
range near the previously reported value (gray area of Fig. \ref{fig:K})
while allowing the semi-amplitude and phase to drift. The central value
near $0\, \kms$ does \textit{not} represent the systemic velocity of the
star, since we have subtracted the mean values of each line from their
nightly values. Our best fit within 5-sigma of the period reported by
G09 is $P = 2.401$ d and $K = 0.05\, \kms$, which is much less than the
putative value of $K = 2.3 \pm 0.1\, \kms$.
\label{fig:limited_phase-fold}}
\end{figure}

\begin{figure}               
\plotone{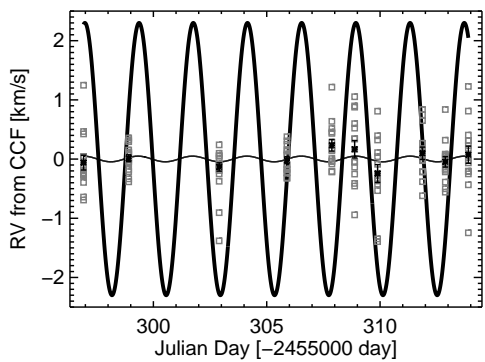}
\caption{The gray boxes show our RV measurements for all
  15 absorption features on each night. The 
  black asterisks show the mean of each night's measurements and the
  associated standard error of the mean.  The thin line shows our best fit
  within a 5-sigma range of $P$ near value reported by G09 (gray area of
  Fig. \ref{fig:K}).  The thick line shows the sinusoid reported
  by G09. As in Fig \ref{fig:limited_phase-fold}, the central value
  near $0\, \kms$ does \textit{not} represent the systemic velocity of the
  star.
\label{fig:full_plus_geier}}
\end{figure}

\begin{figure}                 
\plotone{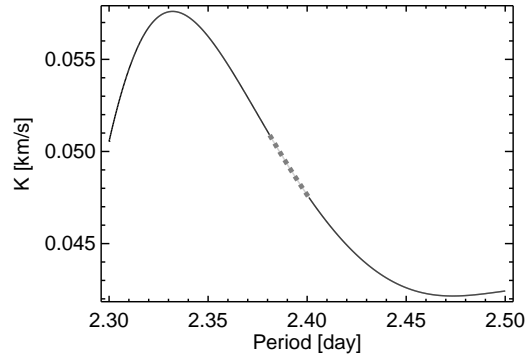}
\caption{Best fit semi-amplitude $K$ as a function of period $P$. The thick,
dotted, gray segment corresponds to a 5-sigma range centered on the period reported
by G09, and includes the fit used in
Fig. \ref{fig:full_plus_geier}. At all points near the putative
$2.391 \pm 0.002$ day period, our measured $K$ is much less than the
value of $2.3 \pm 0.1 \kms$ reported by G09.
\label{fig:K}}
\end{figure}

Our best fit results are period $P=2.401 \, \mathrm{days}$ and RV
semi-amplitude $K = 0.05 \, \kms$ (Fig. \ref{fig:limited_phase-fold}).
This semi-amplitude is much smaller than the claim of $2.3 \pm 0.1 \,
\kms$ by G09, and the difference lies well outside the uncertainties of
the two measurements.  We also fitted our data to a sinusoid with exactly
the semi-amplitude and period reported by G09, but with unknown phase
and zero point offset, using the IDL routine \texttt{mpfitfun}
\citep{Markwardt09}. Figure \ref{fig:full_plus_geier} demonstrates the
discrepancy between our findings and those of G09.  Broadening our
search beyond a 5-sigma window around the putative period, we find that other
best fit values for $K$ for all similarly short periods are below our RV errors
 (Fig. \ref{fig:K}, Table \ref{tab:rv}) indicating that they are
consistent with $0 \, \kms$.

Because some pixel groups contain weaker lines than other groups, we repeat our
analysis for subsets of the groups that omit the poorest features (e.g. omit groups 9 and 12
in Table \ref{tab:line}).  Our results do not change significantly for multiple subsets,
indicating that our exact selection of lines is not important.
Furthermore, as Figure \ref{fig:full_plus_geier} indicates, since the absolute value
of all the RVs in our pixel groups are well underneath the semi-amplitude found by G09,
it is unlikely that any combination of pixel groups will provide a similar fit to that found by G09.

We have no reason to suspect the existence of planets at other periods,
and our experiment was designed only to confirm or disconfirm this
particular claim.  Nonetheless, we have thoroughly searched our data for
any periodicity to determine the parameter space for planets that is
excluded by our data.  We restrict our search to periods between 0.2--28
days, limited both by a semimajor axis $\gtrsim 1$ stellar radius as well as the
longest period easily detectable with our data span.  We further
restrict our search to circular orbits.

The observational cadence of the HET creates ``blind spots'' in our
coverage near the harmonics of 1 sidereal day.  We have excluded these
regions from the following analysis (we define such a ``blind spot'' as
any period for which we have no data points for a stretch of phase
larger than 144 degrees, or 0.4 cycles, and for some
higher harmonics where the data cluster tightly at two phases
180 degrees apart).

\begin{figure}                 
\plotone{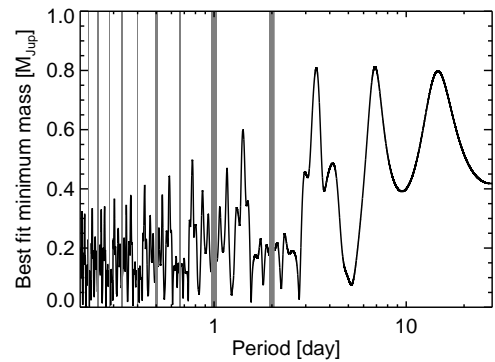}
\caption{Full period-minimum mass diagram for all periods probed in this
work.  ``Blind spots'' created by the cadence imposed by the fixed
elevation of the HET are shaded in gray; in these regions our phase
coverage is too poor to put meaningful constraints on the presence of RV
companions.  Minimum masses correspond to the $M\sin{i}$ values for
planets in circular orbits at the given period with the amplitude of the
best-fit sinusoid to our data at each period.  \label{fig:periodogram}}
\end{figure}

We determined the semi-amplitudes of the best fit sinusoids for a
well-sampled set of periods between 0.2--28 days.  The
largest semi-amplitudes we find are near $0.17 \, \kms$, which we
find at a variety of periods across our search space.  If we adopt the upper limit of the
range of masses of HD 149382 reported by G09 of $0.53 \, \Msun$, then
we can translate these semi-amplitudes into upper limits on the minimum companion
masses ($M\sin{i}$) at each trial period.  We find no best-fit
companions with $M\sin{i}$ in excess of $0.8 \, \Mjup$, which makes the existence
of any close-in giant planet with $M\sin{i} \gtrsim 1 \, \Mjup$
highly unlikely.  Figure~\ref{fig:periodogram}
illustrates our upper limits and blind spots.  In principle, a more
massive companion could exist in a face-on 
orbit, or in one of our period blind spots (or perhaps in a highly
eccentric orbit), but this remaining parameter space is extremely
narrow.

Our results agree with those of \citet{jacobs+2011}, who also find no
massive planets out to $P=50$ days through an analysis of four
\ion{He}{1} lines, although our constraints on the RV variability of HD
149382 are over four times more stringent.

\section{Discussion}   

There are several claims of substellar companions in orbit around hot
subdwarf stars, summarized in \citet{Schuh10}.  With the exception of
HD 149382 and SDSS J08205+0008 (discussed below),
these claims are all based on the interpretation of residuals in an
$O-C$ diagram, in which observed (``$O$'') moments of eclipse or
pulsation maximum are compared with a calculated (``$C$'') ephemeris. A
quasi-sinusoidal pattern of residuals is interpreted in terms of a
light-time effect across the orbit of the subdwarf.  For the eclipsing
systems, the inferred ``planet'' is circumbinary; for the case of
pulsations \citep[V391 Peg:][]{Silvotti07}, the sdB star is otherwise
thought to be single.  Except for HD 149382 and CS 1246 (see below), all
of the claimed orbital periods for substellar companions to sdB stars
are long, in the range of 3 to 16 years.  Nearly always, a parabolic
term is fitted to the $O-C$ residuals along with the orbital sinusoid.
The parabola represents either an evolving pulsation period or an
evolving orbital period of the eclipsing binary, as angular momentum is
transferred by tides from the orbit to the binary companion, which
undergoes ``magnetic braking''.  The relatively short observational
timespans, with the additional issue of correlations between model
parameters describing the parabolic term and the light-time orbit,
conspire to make these claims of substellar companions quite fragile ---
new observations tend to destroy the previous solution. \citep[See][for
examples of broken timing orbits in the case of eclipsing binaries with
white dwarf primaries.]{Parsons10} Quasi-cyclical variation in $O-C$
residuals for binaries may have causes other than orbital motion
involving a third body, such as the ``Applegate mechanism''
\citep{Applegate92}, in which mass quadrupole variations induced by
stellar activity cycles alter the effective Keplerian mass of the
binary.

The only timing-based ($O-C$) detection of a companion to an sdB star
that has been confirmed by an RV study is that of CS 1246
\citep{barlow+2011,BarlowI}. Here the orbital
period is 14.1 days, but the minimum mass of the companion is $M
\sin i \approx 0.13 \, \Msun$, above the substellar limit.

Some of the claimed substellar companions in long-period orbits may be
corroborated with continuing observations, and others will surely be
found.  It is unlikely, however, that these large orbits tell us
anything directly about the formation process of the sdB stars
themselves.  (\citet{Schuh10} does discuss the possibility that large orbits
may be indicators for the previous existence of inner planets, destroyed in a
CE process.) It is also important to bear in mind that only
a lower limit $M \sin i$ to the companion mass $M$ is found from
these timing solutions.  

Short-period orbits of substellar companions around sdB stars are of
more interest for elucidating the prior history of the sdB star.  In
addition to the idea that a planet may assist in
removing the hydrogen envelope from the proto-sdB stars \citep{soker1998},
there is the possibility that ``second-generation'' planets may form in the
debris left from the merger of two helium white dwarfs \citep[the ``helium
planet'' scenario of][]{Silvotti08}.  Note however, that since a merger
between a helium white dwarf and a low-mass hydrogen-burning star may
also create an sdB star \citep{ClausenWade11}, a resulting
second-generation planet need not be of exotic composition.

The formation of sdB stars through envelope stripping by planets is an
intriguing hypothesis, but in light of the fact that the claimed planet
HD 149382 $b$ does not exist, it currently has little observational
support.  To date the only such companion known is that in the eclipsing
sdB+brown dwarf system SDSS J08205+0008, found by \citet{geier+2011}.  The orbital
period is $0.096 \, \mathrm{days}$, and the estimated mass for the strongly irradiated
companion, for reasonable estimates of the mass of the sdB star, is in
the range $0.045 - 0.068 \, \Msun$, slightly below the
hydrogen-burning limit.  This pushes the lower mass limit for companions
in short-period sdB systems to somewhat lower masses than before \citep[see][for
discussion of HW Vir systems]{geier+2011}, but hardly into the planetary
realm. Given the potential importance of significantly substellar companions to the understanding
of the extremes of binary star evolutionary channels as well as the formation of
second-generation planets, searches for such
companions to sdB stars in short-period orbits should continue.

Meanwhile, the putative planet orbiting HD 149382 has already been put
to use by \citet{DeMarco11} in revising the ``$\alpha$ formalism'' for CE
binary evolution. Those authors make the CE efficiency parameter
$\alpha$ a strongly decreasing function of mass ratio.  This conclusion
relies strongly on the large value of $\alpha \sim 33$ inferred for the
putative HD 149382 ``planetary'' system.  \citet{DeMarco11} recognized HD 149382 as
an outlier, and presented parametrized forms for $\alpha$ derived with
and without including it in the sample of studied systems.  With our
demonstration that this system does not exist (at least, as specified by G09), the
weaker dependence of $\alpha$ on mass ratio is favored.

\section{Summary}

Our findings conclusively refute the \citet{geier+2009}
claim of a substellar object orbiting the hot subdwarf star HD 149382.  The orbital solution
found by \citeauthor{geier+2009} is strongly inconsistent with our RV measurements.  Our best fit
results near the putative planet's best period are $P=2.401 \,
\mathrm{days}$ and $K = 0.05 \, \kms$, which is consistent
with $0 \, \kms$ given the uncertainties in our measurements.  We can further
confidently rule out planets with $M\sin{i} > 1 \, \Mjup$ in
circular orbits with periods $< 28$ days, except for a few
special cases arising from the cadence of our observations. 

We argue that the evolutionary origins of this hot subdwarf cannot
involve an extant, close-in companion of giant planet mass or greater.

\acknowledgements

This work was partially supported by funding from the Center for
Exoplanets and Habitable Worlds.  The Center for Exoplanets and
Habitable Worlds is supported by the Pennsylvania State University, the
Eberly College of Science, and the Pennsylvania Space Grant Consortium.
R.W. gratefully acknowledges support from National Science
Foundation grant AST-0908642.

The Hobby-Eberly Telescope (HET) is a joint project of the University of Texas at Austin, the Pennsylvania State University, Stanford University, Ludwig-Maximillians-Universit\"at M\"unchen, and Georg-August-Universit\"at G\"ottingen. The HET is named in honor of its principal benefactors, William P. Hobby and Robert E. Eberly.


\end{document}